\documentclass[aps,prl,twocolumn,showpacs,tightenlines,]{revtex4}
\usepackage{graphicx}
\usepackage{epsfig}
\begin{document}

\def \cT {{\cal T}}
\def \cI {{\cal I}}
\def \cf {{\cal f}}
\def \cG {{\cal G}}
\def \cD {{\cal D}}
\def \cU {{\cal U}}
\def \cV {{\cal V}}
\def \cF {{\cal F}}
\def \cT {{\cal T}}
\def \cH {{\cal H}}
\def \cA {{\cal A}}
\def \cL {{\cal L}}
\def \cR {{\cal R}}
\def \cN {{\cal N}}
\def \cC {{\cal C}}
\def \cS {{\cal S}}
\def \cP {{\cal P}}
\def \cE {{\cal E}}
\def \cM {{\cal M}}

\title{Electron transport in interacting hybrid mesoscopic systems}
\author{Z. Y. Zeng$^1$, Baowen Li$^1$ and F. Claro$^2$}
\affiliation{$^1$Department of Physics, National University of
Singapore, 117542, Singapore\\
$^2$Facultad de F\'isica, Pontificia Universidad Cat\'olica de
Chile, Casilla 306, Santiago 22, Chile }

\date{\today}

\begin{abstract}
A unified theory  for the current through a mesoscopic region of
interacting electrons connected to two leads which can be either
ferromagnet or superconductor is presented, yielding
Meir-Wingreen-type formulas when applied to specific
circumstances. In such a formulation, the requirement of gauge
invariance is satisfied automatically. Moreover,  one can judge
unambiguously  what quantities can be measured in the transport
experiment.

\end{abstract}


\maketitle

Mesoscopic electron transport has received an increasing attention
both theoretically and experimentally in last
decade\cite{Kouwenhoven}. In mesoscopic or nanoscale systems the
wave nature of electrons becomes apparent and the transport
process is coherent. The Landauer-B\"uttiker
formula\cite{Landauer}, which encodes the current in the local
properties of the interacting mesoscopic region and the
equilibrium distribution functions of the noninteracting electron
reservoirs, enhances our understanding of mesoscopic electron
transport and has been applied successful in many
fields\cite{Datta}. In 1992 Meir and Wingreen \cite{Meir}
presented a formulation for electron transporting through a small
confined region (quantum dot, QD) where the electron-electron
interaction is important, and recovered the form of
Landauer-B\"uttiker formula in the noninteracting case.

Recent advances in nanofabrication and material growth
technologies make it possible to realize various kinds of hybrid
mesoscopic structures\cite{Poirier, Post, Morpurgo,
Tuominen,Eiles,Upadhyay, Lawrence,Gueron},
 of which the building blocks are
normal metals (N), ferromagnets (F) and superconductors (S). It is
known that transport in the presence of a ferromagnet and a
superconductor will be strongly related to the spin polarization
of the ferromagnet and the Andreev reflection at the boundary of
the superconductor \cite{Jong, Prinz}. The co-existence of two
ferromagnets or two superconductors is revealed to display
spin-valve effect \cite{Slonczewski} or Josephson
effect\cite{Tinkham}. When an interacting normal metal is
connected with bulk ferromagnet(s) and/or superconductor(s), it is
expected that the interplay among the electron-electron
interaction, spin imbalance, Andreev reflection would induce more
interesting or even more surprising features in mesoscopic
electronic transport. Previous theoretical investigations on the
transport properties of specific mesoscopic hybrid structures,
such as N-QD-S\cite{Fazio}, S-QD-S\cite{Yeyati},
F-QD-F\cite{Sergueev}, F-QD-S \cite{Zhu} etc., make some intuitive
presumptions and hence lack mathematical rigidity, which will be
discussed below in detail .

In this paper, we provide a scheme to treat the transport problem
in an interacting hybrid mesoscopic structure in a unified way by
using the Keldysh formalism\cite{Meir, Jauho}. It is shown that
gauge invariance  can be satisfied automatically. It is also shown
that what physical quantities can be measured in experiment.

We start with the Hamiltonian
\begin{equation}
\label{total-Hamiltonian}
\cH=\cH_{\cL}+\cH_{\cR}+\cH_{\cC}+\cH_{\cT}, \nonumber
\end{equation}
where  $ \cH_{\cC} =\sum_{n\sigma }(\varepsilon _{n\sigma
}-\mu_\cC)\psi_{dn\sigma }^{\dagger }\psi_{dn\sigma }+
 \cH_{int}\left(\{\psi_{dn\sigma
}^{\dagger}\},\{\psi_{dn\sigma }\}\right)$ is the Hamiltonian for
the central interaction region,  $\cH_\cL$ ($\cH_\cR$) for the
left (right) lead can be either the {\it Stoner} model
\cite{Slonczewski} characterized by an exchange magnetization $h$
with polar angle $\theta$ or the {\it BCS} Hamiltonian
\cite{Tinkham} with order parameter $\Delta=|\Delta|e^{i\varphi}$:
 $
 \cH^{(F)}_\gamma =\sum_{k\sigma
}[\varepsilon _{\gamma k\sigma }+sign(\sigma) h_{\gamma} \cos
\theta_{\gamma f}-\mu_\gamma ]f_{\gamma k\sigma }^{\dagger
}f_{\gamma k\sigma } +\sum_{k\sigma }h_{\gamma}\sin \theta_{\gamma
f} f_{\gamma k\sigma }^{\dagger}f_{\gamma k\stackrel{-}{\sigma }},
$ or $
 \cH^{(S)}_\gamma =\sum_{k\sigma
}(\varepsilon_{\gamma k\sigma }-\mu_\gamma)s_{\gamma k\sigma
}^{\dagger }s_{\gamma k\sigma }+  \sum_{k}\left[ \Delta
^{*}_\gamma s_{\gamma k\uparrow }^{\dagger }s_{\gamma -k\downarrow
}^{\dagger }+\Delta_\gamma s_{\gamma -k\downarrow } s_{\gamma
k\uparrow }\right]$, tunneling Hamiltonian can be written as $
\cH_{\cT}^{\gamma {(F)}} =\sum_{kn;\sigma }\left[ V^{\gamma
f}_{kn;\sigma }f_{\gamma k\sigma }^{\dagger }\psi_{dn\sigma
}+H.c.\right]$  or $ \cH_{\cT}^{\gamma {(S)}} =\sum_{kn;\sigma
}\left[ V^{\gamma s}_{kn;\sigma }s_{\gamma k\sigma }^{\dagger
}\psi_{dn\sigma }+H.c.\right].$ Here $\gamma=\cL, \cR$,
$\mu_{\gamma/\cC}$  is the chemical potential of the corresponding
part, and hereafter the notations $\sigma=\uparrow, \downarrow$
and $\sigma=\pm$ are used interchangeably.

To see the tunneling processes more clearly, and even more
importantly, to facilitate the analysis of the gauge invariance,
and the simplification of the general current formula
(\ref{eq:current-final}) to the form of specific systems, we first
transform the {\it Stoner} and {\it BCS} Hamiltonian and the
tunneling Hamiltonian by the following Bogoliubov transformations
\begin{eqnarray}
f_{\gamma k\sigma}=\cos (\theta_{\gamma f}/2) \psi_{\gamma
fk\sigma}-sgn(\sigma) \sin(\theta_{\gamma f}/2) \psi_{\gamma fk
\stackrel{-}{\sigma}}, \nonumber\\
e^{-i\varphi_\gamma/2}s_{\gamma k\sigma}=\cos \theta_{\gamma sk'}
\psi_{\gamma sk'\sigma}+sgn(\sigma) \sin \theta_{\gamma s}\cP
\psi_{\gamma sk'\stackrel{-}\sigma}^{\dagger},\nonumber
\end{eqnarray}
where $\theta_{\gamma sk'}=\arctan[(\varepsilon_{\gamma k\sigma}+
\sqrt{\varepsilon_{\gamma
k\sigma}^2+\Delta^2_\gamma})/(\varepsilon_{\gamma k\sigma}-
\sqrt{\varepsilon_{\gamma k\sigma}^2+\Delta^2_\gamma})]^{1/2}$ and
$\cP$ ($\cP^{\dagger}$) is the pair destruction (creation)
operator guaranteeing the particle conservation, transforming  a
state of a given $N$ particles into that with $(N+2)$/$(N-2)$
particles,  i.e.,
$\cP^\dagger/\cP|N\rangle=|N+2\rangle/|N-2\rangle$ , thus make
sense the abnormal off-diagonal Green's functions consists of two
creation or destruction particle operators. One finds that the
lead Hamiltonian is diagonalized after the above Bogoliubov
transformations.

In order to treat  the ferromagnet and superconductor on the same
footing, we introduce a 4-dimensional Nambu-spinor space, denoted
by ${\bf \Psi}_\alpha= \left(\matrix{ \psi_{\alpha\uparrow
}^{\dagger } & \psi_{\alpha \downarrow } & \psi_{\alpha\downarrow
}^{\dagger } & \psi_{\alpha\uparrow} \cr } \right)^\dagger$ and
the Green's function in the Keldysh formalism \cite{Jauho} ${\bf
G}_{\alpha,\beta}(t_1,t_2)=i \langle T_C({\bf \Psi}_\alpha(t_1)
\otimes {\bf \Psi}_\beta^\dagger(t_2))\rangle$,
 where $T_c$ is the contour-order operator, including
 ${\bf
G}_{\alpha, \beta}^{r/a}(t_1,t_2)= \mp i \vartheta(\pm t_1 \mp
t_2)\langle\{{\bf \Psi}_{\alpha }(t_1),{\bf \Psi}_{\beta
}^\dagger(t_2)\}_+ \rangle$, and  $ {\bf G}_{\alpha , \beta
}^{</>}(t_1,t_2)=\pm i\langle{\bf \Psi}_{\beta }^\dagger(t_2)/{\bf
\Psi}_{\alpha }(t_1) \otimes {\bf \Psi}_{\alpha }(t_1) /{\bf
\Psi}_{\beta }^\dagger(t_2)\rangle$. The tunneling Hamiltonian in
such a representation takes the form
\begin{eqnarray}
\cH^{\gamma {(F)}}_\cT &=&\sum_{kn}\big({\bf \Psi}_{\gamma
fk}^\dagger {\bf
V}^{\gamma f}_{kn}(t){\bf \Psi}_{dn}+H.c.\big), \\
 \cH^{\gamma {(S)}}_\cT&=&\sum_{kn}\big({\bf \Psi}_{\gamma sk'}^\dagger {\bf
V}^{\gamma s}_{kn}(t){\bf \Psi}_{dn}+H.c.\big),
 \end{eqnarray}
 where ${\bf
V}^{\gamma f}_{kn}(t)={\bf R}^f(\frac{\theta_{\gamma f}}{2}){\bf
V}^{\gamma f}_{kn}{\bf P}(\mu_{\gamma \cC} t)$,
 ${\bf
V}^{\gamma s}_{kn}(t)={\bf R}^s(\theta_{\gamma sk'}){\bf
V}^{\gamma s}_{kn}{\bf P}(\mu_{\gamma \cC}
t+\frac{\varphi_\gamma}{2})$, with
$\mu_{\gamma\cC}=\mu_\gamma-\mu_\cC$,
\begin{eqnarray}
{\bf V}^{\gamma f/s}_{kn}
 =\left (\matrix{ V^{\gamma f/s}_{kn} &
 0 & 0
  & 0 \cr
 0 &-V^{\gamma f/s*}_{kn}& 0 &
  0\cr
 0 & 0
 &V^{\gamma f/s}_{kn} & 0 \cr
 0 & 0 & 0 &
  -V^{\gamma f/s*}_{kn} \cr }\right),\nonumber
  \end{eqnarray}
and the  unitary {\it rotation} and {\it phase}
operators(matrices) are given by
\begin{eqnarray*}
{\bf R}^{ f}(x)
 &=&\left (\matrix{ \cos x &
 0 & \sin x
  & 0 \cr
 0 &\cos x& 0 &
  -\sin x\cr
 -\sin x& 0
 &\cos x& 0 \cr
 0 & \sin x & 0 &
  \cos x\cr }\right), \\
{\bf R}^{ s}(x)
 &=&\left (\matrix{ \cos x &
  -\cP\sin x &0
  & 0 \cr
 \cP^*\sin x  &
 \cos x& 0 & 0\cr
 0& 0&\cos x &\cP\sin x \cr
 0 &0 & -\cP^*\sin x  &
\cos x\cr }\right), \\
 {\bf P}(x) &=& \left(\matrix{ e^{ix/\hbar} & 0 & 0 &0\cr
 0& e^{-ix/\hbar}&0&0\cr
 0 & 0&e^{ix/\hbar} & 0 \cr
 0 & 0 & 0& e^{-ix/\hbar}\cr }\right).
\end{eqnarray*}
The chemical potential $\mu_{\gamma/\cC}$ is incorporated into the
{\it phase} matrix ${\bf P}$ and the ferromagnetism and
superconductivity are reflected in the corresponding {\it
rotation} matrices ${\bf R}^{ f}$ and ${\bf R}^{ s}$. The
tunneling Hamiltonian now represents the explicit physical
processes in the {\it semiconductor model}\cite{Tinkham}: an
electron of spin $\sigma$ in the central regime can tunnel into
either the spin $\sigma$ band or $\stackrel{-}{\sigma}$ band of
the ferromagnetic lead, or tunnel into a spin $\sigma$ state or
condensate into an electron pair with a hole state of opposite
spin being created; and vice versa. The total probability of the
two tunneling processes into the same lead is $\cos^2x+\sin^2x=1$.
As shown below, these {\it rotation} and {\it phase} matrices are
very useful in our analysis of gauge invariance,  and even more
importantly, the simplification of the formulas.

 The current from
the left lead into the interacting region is\cite{Meir}
\begin{eqnarray}
\label{eq:current-left}
\cI_{\cL}(t)&=&-e\langle \dot{N}_\cL\rangle \nonumber\\
&=& \frac{2e}{\hbar}Re\sum\limits_{nk}^{i=1,3}\Big({\bf V}^{\gamma
f/s\dagger }_{kn}(t){\bf G}_{\gamma
f/sk,dn}^<(t,t)\Big)_{ii}\nonumber\\
 &=&
\frac{2e}{\hbar}Re\sum\limits^{i=1,3}_{nm}\int_{-\infty}^{t}dt_1
\Big({\bf \Sigma}^r_{\cL f/s; nm}(t,t_1){\bf
G}_{dm,dn}^{<}(t_1,t)\nonumber\\
&&\hspace{1.5cm} +{\bf \Sigma}^<_{\cL f/s; nm}(t,t_1){\bf
G}_{dm,dn}^{a}(t_1,t)\Big)_{ii},
\end{eqnarray}
where the self-energy matrices after converting the sum $\sum_{k}$
into an integral $\int d\varepsilon_{k} \rho^{\gamma
f/s}_{\sigma/N}(\varepsilon_{k})$(where $\rho^{\gamma
f/s}_{\sigma/N}$  is the spin-dependent/normal density of states
of the feromagnet/superconductor) are
\begin{eqnarray}
\label{eq:arselfenergy-f}
 {\bf \Sigma}^{r/a}_{\gamma f;mn}(t_1,t_2)
&=&\mp \frac{i}{2}\int \frac{d\varepsilon}{2\pi}
e^{-i\varepsilon(t_1-t_2)/\hbar}{\bf R}^{f\dagger}(\frac{\theta_{\gamma  f}}{2}) \nonumber\\
 &&\hspace{1.5cm}{\bf
\Gamma}^{\gamma f}_{mn}(\varepsilon\mp\mu_{\gamma \cC}){\bf
R}^f(\frac{\theta_{\gamma  f}}{2}), \\
\label{eq:arselfenergy-s}
 {\bf \Sigma}_{\gamma
s;mn}^{r/a}(t_1,t_2)
 &=&\mp \frac{i}{2}\int \frac{d\varepsilon}{2\pi}
e^{-i\varepsilon(t_1-t_2)/\hbar}{\bf P}^\dagger(\mu_{\gamma\cC}t_1+\frac{\varphi_\gamma}{2}) \nonumber\\
&& \hspace{0cm}{\bf \Gamma}^{\gamma
s}_{\varrho/\varrho^*;mn}(\varepsilon\mp \mu_{\gamma \cC}){\bf
P}(\mu_{\gamma \cC}t_1+\frac{\varphi_\gamma}{2}),\\
\label{eq:lgselfenergy-f}
{\bf \Sigma}^{</>}_{\gamma
f;nm}(t_1,t_2) &=&i\int \frac{d\varepsilon}{2\pi}
e^{-i\varepsilon(t_1-t_2)/\hbar}{\bf
R}^{f\dagger}(\frac{\theta_{\gamma  f}}{2})  \nonumber\\
&&\hspace{-2.2cm} {\bf \Gamma}^{\gamma f}_{nm}(\varepsilon\mp
\mu_{\gamma \cC})
 {\bf R}^f(\frac{\theta_{\gamma  f}}{2}) [{\bf
f}_{\gamma}(\varepsilon\mp \mu_{\gamma \cC} )-\frac12{\bf 1}\pm
\frac12{\bf 1}], \\
\label{eq:lgselfenergy-s}
{\bf \Sigma}^{</>}_{\gamma
s;nm}(t_1,t_2) &=& i \int \frac{d\varepsilon}{2\pi}
e^{-i\varepsilon(t_1-t_2)/\hbar}{\bf P}^\dagger(\mu_{\gamma \cC}
t_1+\frac{\varphi_\gamma}{2})\nonumber
\\&&\hspace{-1cm}{\bf \Gamma}^{\gamma
s}_{\rho;nm}(\varepsilon\mp\mu_{\gamma \cC})[{\bf
f}_{\gamma}(\varepsilon\mp\mu_{\gamma \cC})-\frac12{\bf 1}\pm
\frac12{\bf 1}]\nonumber\\&&\hspace{2cm} {\bf P}(\mu_{\gamma \cC}
t_1+\frac{\varphi_\gamma}{2}),
\end{eqnarray}
with (subscripts $mn$ is omitted in the matrices)
\begin{eqnarray*}
{\bf f}_{\gamma}(\varepsilon\mp c)=\nonumber \\
 &&\hspace{-1cm}
 \left (\matrix{ f(\varepsilon-c) & 0 & 0
  & 0 \cr
 0 &f(\varepsilon+c) & 0 & 0 \cr
 0 & 0&f(\varepsilon-c)  & 0 \cr
 0 & 0 & 0 & f(\varepsilon+c)  \cr}\right),\\
{\bf \Gamma}^{\gamma f}(\varepsilon\mp
c)&=&\nonumber \\
 &&\hspace{-2.3cm}\left (\matrix{ \Gamma_{\uparrow}^{\gamma
f}(\varepsilon-c) & 0 & 0& 0 \cr
 0 &\Gamma_{\downarrow}^{\gamma
f}(\varepsilon+c)& 0 & 0\cr
 0 &
0&\Gamma_{\downarrow}^{\gamma f}(\varepsilon-c)& 0 \cr
 0 & 0 & 0 & \Gamma_{\uparrow}^{\gamma
f}(\varepsilon+c) \cr }\right),\nonumber\\
&& \Gamma_{\sigma;mn}^{\gamma f}(\varepsilon)=2\pi \rho^{\gamma
f}_{\sigma}(\varepsilon)V^{\gamma f\dagger}_{km}V^{\gamma f}_{kn},
\end{eqnarray*}
and ($x=\varrho, \varrho^*, \rho$)
\begin{eqnarray*}
 {\bf \Gamma}^{\gamma
s}_{x}(\varepsilon\mp c)&=&\Gamma^{\gamma s}\nonumber\\
&&\hspace{-2.2cm} \left(\matrix{x^{\gamma s}(\varepsilon-c)
  & -\frac{|\Delta_\gamma|}{\varepsilon+c}x^{\gamma
s}(\varepsilon+c) \cr
-\frac{|\Delta_\gamma|}{\varepsilon-c}x^{\gamma s}(\varepsilon-c)&
x^{\gamma s}(\varepsilon+c) \cr
 0 & 0\cr 0&0\cr}\right.\nonumber\\
 && \hspace{0.1cm} \left.\matrix{ 0 & 0 \cr
  0 & 0 \cr
 x^{\gamma
s}(\varepsilon-c)&\frac{ |\Delta_\gamma|}{\varepsilon+c}x^{\gamma
s}(\varepsilon+c)\cr
 \frac{
 |\Delta_\gamma|}{\varepsilon-c}x^{\gamma
s}(\varepsilon-c)&x^{\gamma s}(\varepsilon+c)
\cr}\right), \nonumber \\
\varrho^{\gamma
s}(\varepsilon)&=&-i\frac{\varepsilon\vartheta(|\Delta_\gamma|-|\varepsilon|)}
{\sqrt{|\Delta_\gamma|^2-\varepsilon^2}}+\frac{|\varepsilon|\vartheta
(|\varepsilon|-|\Delta_\gamma|)}
{\sqrt{\varepsilon^2-|\Delta_\gamma|^2}},\nonumber\\
\Gamma^{\gamma s}_{mn}&=&2\pi \rho^{\gamma s}_N(0) V^{\gamma
f\dagger}_{km}V^{\gamma f}_{kn}.
\end{eqnarray*}
Here $f(x)=1/(1+e^x/k_BT)$ is the Fermi distribution function. The
quasi-particle density of states of the {\it BCS} superconductor
is $ \rho^{\gamma s}(\varepsilon)= Re\{\varrho^{\gamma
s}(\varepsilon)\}$.

The current flowing from the right lead $\cI_\cR$ can be obtained
in a similar way.  In the steady transport problem, the current is
uniform if no charge piles up in the central region, that is,
$\cI_\cL=-\cI_\cR$. Symmetrizing Eq. (\ref{eq:current-left}) one
finds
\begin{eqnarray}
\label{eq:current-final}
 \cI(t)&=&
\frac{e}{\hbar}Re\sum\limits^{i=1,3}\int\limits_{-\infty}^{t}dt_1Tr\Big\{
\big[{\bf \Sigma}^r_{\cL f/s}(t,t_1)-{\bf \Sigma}^r_{\cR
f/s}(t,t_1)\big] \nonumber\\ && \hspace{-1.2cm} {\bf
G}_{d,d}^{<}(t_1,t)+\big[ {\bf \Sigma}^<_{\cL f/s}(t,t_1)-{\bf
\Sigma}^<_{\cR f/s}(t,t_1)\big] {\bf
G}_{d,d}^{a}(t_1,t)\Big\}_{ii},\nonumber\\
\end{eqnarray}
where the trace is over the level indices in the central region.
Eq. (\ref{eq:current-final}) with the self-energy matrices
(\ref{eq:arselfenergy-f},\ref{eq:arselfenergy-s},\ref{eq:lgselfenergy-f},\ref{eq:lgselfenergy-s})
is the central result of this work, it expresses the current
through an interacting region in terms of the local properties of
such region (${\bf G}^{r,a/<}_{d,d}$) and the equilibrium
distribution functions (${\bf \Sigma}^{r,a/<}_{\gamma f/s}$) of
the attached leads, as in the work of Meir and Wingreen
\cite{Meir}. It can be applied to many types of hybrid mesoscopic
structures even in the non-equilibrium situation, allowing various
kinds of interactions in the central region. Notice that, the
current is generally time-independent, except the case of two
superconducting leads with nonzero bias, and that the full Green's
functions ${\bf G}^{r,a/<}_{d,d}$ should be evaluated with the
consideration of the tunnelling between the interacting region and
the leads. The retarded/advanced Green's functions ${\bf
G}^{r,a}_{d,d}$ can be calculated in several approaches, such as
the equation of motion formalism\cite{Wingreen}, interpolative
method\cite{Rodero} and the NCA technique\cite{Bickers}. While the
lesser one ${\bf G}^{<}_{d,d}$ can be obtained from the Keldysh's
equation based on Ng's Ansatz\cite{Ng}.
 With the help of the unitary property of the
{\it phase} operator ${\bf P}$, it is not difficult to find that
Eq. (\ref{eq:current-final}) is gauge invariant, namely, $\cI(t)$
remains unchanged under a global energy shift. When the bias
$V=(\mu_\cL-\mu_\cR)/e$ becomes zero, one can readily find that
$\cI(t)=0$ except the case of two superconducting leads with
different superconducting phases, due to the coherent transport of
quasi-particle pairs\cite{Zeng}.

In the following, we apply this generalized result Eq.
(\ref{eq:current-final}) to the specific structures we are
interested in. When the two leads are both ferromagnetic, after a
{\it phase} and a {\it rotation} transformation,  Eq.
(\ref{eq:current-final}) reduces to
\begin{eqnarray}
\label{eq:current-fnf}
 \cI_{fnf}&=&\frac{ie}{2\hbar}\sum\limits^{i=1,3}\int\frac{d\varepsilon}{2\pi}
Tr\Big\{\Big(\big[\hat{{\bf \Gamma}}^{\cL f}(\varepsilon\mp
eV)-{\bf \Gamma}^{\cR f}(\varepsilon)\big]
\nonumber \\
&& \hspace{0.2cm} \widehat{{\bf G}}_{d,d}^{<}(\varepsilon)+
\big[\hat{{\bf \Gamma}}^{\cL f}(\varepsilon\mp eV){\bf
f}_\cL(\varepsilon\mp
eV)-\nonumber\\
&& \hspace{0.5cm}{\bf \Gamma}^{\cR f}(\varepsilon){\bf
f}_\cR(\varepsilon)\big] \big[\widehat{{\bf
G}}_{d,d}^{r}(\varepsilon)-\widehat{{\bf
G}}_{d,d}^{a}(\varepsilon)\big]\Big)_{ii} \Big\},
\end{eqnarray}
where $\hat {\bf \Gamma}^{\cL f}={\bf R}^{f\dagger}(\frac{\theta_{
f}}{2}) {\bf \Gamma}^{\cL f}{\bf R}^f(\frac{\theta_{f}}{2})$, $
\theta_f=\theta_{\cL f}-\theta_{\cR f}$, and
\begin{eqnarray*}
\widehat {\bf G}^{r,a/<}_{d,d}(\varepsilon)&=&\int
d(t-t')e^{i\varepsilon(t-t')/\hbar}{\bf P}(\mu_{\cR \cC} t){\bf
R}^f(\frac{\theta_{\cR f}}{2}) \nonumber\\
&&\hspace{1cm} {\bf G}^{r,a/<}_{d,d}(t,t'){\bf
R}^{f\dagger}(\frac{\theta_{\cR f}}{2}){\bf P}^\dagger(\mu_{\cR
\cC} t').
\end{eqnarray*}
One sees that Eq. (\ref{eq:current-fnf}) is formally the same as
the Meir-Wingree formula \cite{Meir} in the normal lead case. The
current is time-independent as one might expect, and just depends
on the relative angle $\theta_f$ between two magnetization
orientations and the bias $V$, the difference between the chemical
potential of the two leads.

The current through an interaction region with a ferromagnetic and
a superconducting lead can be derived similarly
\begin{eqnarray}
\label{eq:current-fns}
 \cI_{fns}&=&\frac{ie}{2\hbar}\sum\limits^{i=1,3}\int\frac{d\varepsilon}{2\pi}
Tr\Big\{\Big(\big[{\bf \Gamma}^{\cL f}(\varepsilon\mp
eV)-{\bf \Gamma}^{\cR s}_\rho(\varepsilon)\big]\nonumber \\
&& \hspace{0.2cm}\overline{{\bf G}}_{d,d}^{<}(\varepsilon)+
\big[{\bf \Gamma}^{\cL f}(\varepsilon\mp eV){\bf
f}_\cL(\varepsilon\mp
eV)-\nonumber\\
&& \hspace{0.5cm}{\bf \Gamma}^{\cR s}_\rho(\varepsilon){\bf
f}_\cR(\varepsilon)\big] \big[\overline{{\bf
G}}_{d,d}^{r}(\varepsilon)-\overline{{\bf
G}}_{d,d}^{a}(\varepsilon)\big]\Big)_{ii} \Big\},
\end{eqnarray}
in which the full Green's functions are
\begin{eqnarray*}
\overline {\bf G}^{r,a/<}_{d,d}(\varepsilon)&=&\int
d(t-t')e^{i\varepsilon(t-t')/\hbar}{\bf P}(\mu_{\cR \cC}
t+\frac{\varphi_\cR}{2}) \nonumber\\
&&\hspace{-1.1cm} {\bf R}^f(\frac{\theta_{\cL f}}{2}){\bf
G}^{r,a/<}_{d,d}(t,t'){\bf R}^{f\dagger}(\frac{\theta_{\cL
f}}{2}){\bf P}^\dagger(\mu_{\cR \cC} t'+\frac{\varphi_\cR}{2}).
\end{eqnarray*}
The current through a generic hybrid mesoscopic structure also
depends on just the bias $V$.  The dependence on the magnetization
orientation of the ferromagnetic lead and the phase of the order
parameter of the superconductor lead is absent after the {\it
rotation} and {\it phase} operations, which are just presumptions
in previous investigations\cite{Fazio, Yeyati, Sergueev,Zhu}.

When the two leads are both superconductors, the situation becomes
complicated. However, we  still obtain an elegant formula in this
case following the same procedure as in the above derivation
\begin{eqnarray}
\label{eq:current-sns}
 \cI_{sns}(t) &=&-\frac{e}{\hbar}\sum\limits^{i=1,3}\int\frac{d\varepsilon}{2\pi}
ImTr\Big\{\Big(\frac12\big[\tilde{{\bf \Gamma}}^{\cL
s}_\varrho(\varepsilon\mp
eV;t)-\nonumber \\
&& \hspace{1cm}{\bf \Gamma}^{\cR
s}_\varrho(\varepsilon)\big]\widetilde{{\bf
G}}_{d,d}^{<}(\varepsilon;t)- \big[\tilde{{\bf \Gamma}}^{\cL
s}_\rho(\varepsilon\mp eV;t)\nonumber\\
&&  \hspace{-0.3cm}{\bf f}_\cL(\varepsilon\mp eV)-{\bf
\Gamma}^{\cR s}_\rho(\varepsilon){\bf
f}_\cR(\varepsilon)\big]\widetilde{{\bf
G}}_{d,d}^{a}(\varepsilon;t) \Big)_{ii}\Big\},
\end{eqnarray}
where
\begin{eqnarray*}
\tilde{{\bf \Gamma}}^{\cL s}_{\varrho/\rho}(\varepsilon\mp
eV;t)&=&{\bf P}^\dagger(eVt+\frac{\varphi_s}{2}){\bf \Gamma}^{\cL
s}_{\varrho/\rho}(\varepsilon\mp eV)\nonumber\\
 && \hspace{3cm} {\bf P}(eVt+\frac{\varphi_s}{2}),\\
\widetilde{{\bf G}}_{d,d}^{r,a/<}(\varepsilon;t)&=&\int
d(t-t')e^{i\varepsilon(t-t')/\hbar}{\bf P}(\mu_{\cR \cC}
t+\frac{\varphi_\cR}{2})\nonumber\\
&& \hspace{1cm}{\bf G}^{r,a/<}(t,t'){\bf P}^\dagger(\mu_{\cR \cC}
t'+\frac{\varphi_\cR}{2}),
\end{eqnarray*}
with $\varphi_s=\varphi_\cL-\varphi_\cR$. Here we have added the
time variable $t$  into  the full Green's function
$\widetilde{{\bf G}}_{d,d}^{r,a/<}(\varepsilon;t)$, other than
$\widehat{{\bf G}}_{d,d}^{r,a/<}(\varepsilon)$. The reason is that
 the full Green's functions  should be calculated
 in the presence of tunneling as well as interactions in the
 central region. In the present case, the $t$-dependence can not
 be avoided in the self-energy matrices (\ref{eq:arselfenergy-s},\ref{eq:lgselfenergy-s}),  while it can be
 removed by a unitary {\it phase} operation in the case of only one
 superconductor lead. The current  through an interacting
 mesoscopic region with two superconductor leads, is generally
 time dependent, as in the case of weak Josephson
 links\cite{Tinkham}. However, in the limiting case of zero bias,
 the current is a time-independent nonzero quantity, as can be
 found from Eq. (\ref{eq:current-sns}).

 The time-dependence of $\widetilde {\bf G}^{r,a/<}_{d,d}$ arises
 from the couplings to the voltage biased two superconducting leads.  One can
 show that $\widetilde {\bf G}^{r,a/<}_{d,d}$ depends only on the
 single time variable $t$  at least within perturbative analysis.
 In fact the Green's function $\widetilde {\bf G}^{r,a/<}_{d,d}$ can be expanded
 in powers
 of the fundamental frequency $\omega_0=2eV/\hbar$, i.e., $\widetilde {\bf
 G}^{r,a/<}_{d,d}(\varepsilon,t)=\sum_m \widetilde {\bf
 G}^{r,a/<}_{d,d}(\varepsilon,\varepsilon+m\omega_0/2)e^{im\omega_0
 t/2}$, which with the expression for the Green's function $\widetilde {\bf G}^{r,a/<}_{d,d}$
 below Eq.(\ref{eq:current-sns}) is
 exactly the form of  the double-energy transformation\cite{Arnold}. And the current can be
 generally expressed as $\cI(t)=\sum_n I_n e^{in\omega_0 t}$.

 The ferromagnetic or superconductor
 lead will be in a normal state, when the magnetization $h$
 or the order parameter $\Delta$ becomes zero. Such an observation
 allows us to study a more broad category of the hybrid structures
 using the above formalism. By defining the unitary operators, we
 prove rigorously that the {\it relative} value between some quantities,
 such as chemical potential, magnetization orientation, and order
 parameter phase etc. can be measured. {\it All these quantities can be
 expressed as the energy-independent arguments of the exponential functions or the
 triangle functions in a unitary matrix},  reflecting some kind of requirement
 of the symmetrical invariance.  As a further example, we recover all the
 known formulas obtained
 in the absence of interactions{\cite{Fazio, Yeyati,
Sergueev,Zhu,Zeng}. This proves the validity and generality of our
formalism in other respects. Taking into consideration the
electron-electron interactions, one expect rich physics to show
up, especially in the Kondo regime.

In summary, we have given a unified formula for the current
through an interacting region with either superconducting or
ferromagnetic leads. The current formula derived satisfies gauge
invariance automatically. Such a current formula can be applied to
an appreciable class of hybrid mesoscopic systems, allowing
arbitrary interactions within the central nanoscale region.

ZZ and BL have been supported in part by the Academic Research
Fund of the National University of Singapore and DSTA of
Singapore. FC
 has been supported by
 C\'atedra Presidencial en Ciencias of Chile and
  FONDECYT 1020829 of Chile.

\end{document}